# Visualization of carrier transport in lateral metal-perovskite-metal structures and its influence on device operation

N. Ganesh[1], A.Z. Ashar[1], Sumukh Purohit[1], K.L. Narasimhan[2] and K.S. Narayan[1]*

[1]*Chemistry and Physics of Materials Unit (CPMU), Jawaharlal Nehru Center for Advanced Scientific Research, Jakkur, Bengaluru, India – 560064*

[2]*Center for Nano Science and Engineering (CENSE), Indian Institute of Science, Bengaluru, India - 560012*

## Abstract

The high performance of hybrid perovskite-based devices is attributed to its excellent bulk-transport properties. However, carrier dynamics, especially at the metal-perovskite interface, and its influence on device operation are not widely understood. This work presents the dominant transport mechanisms in methylammonium lead iodide (MAPbI$_3$) perovskite-based asymmetric metal-electrode lateral devices. The device operation is studied with inter-electrode lengths varying from 4 μm to 120 μm. Device characteristics indicate distinct ohmic and space-charge limited current (SCLC) regimes that are controlled by the inter-electrode length and applied bias. The electric-potential mapping using Kelvin-Probe microscopy across the device indicates minimal ion-screening effects and the presence of a transport barrier at the metal-MAPbI$_3$ junction. Further, photocurrent imaging of the channel using near-field excitation-scanning microscopy reveals dominant recombination and charge-separation zones. These lateral devices exhibit photodetector characteristics with a responsivity of about 51 mA/W in self-powered mode and 5.2 A/W at 5 V bias, in short-channel devices (4 μm). The low device capacitance enables a fast light-switching response of ~12 ns.

A. Introduction

Hybrid organic-inorganic perovskites (HOIP) have evolved as attractive materials for optoelectronic applications. These materials have exhibited high performance in solar cells, photodetectors, LEDs, neuromorphic devices, and lasing applications [1-7]. The excellent performance of these devices is attributed to their outstanding bulk transport properties such as long diffusion length, extended carrier lifetimes, high absorption coefficient, and mobility [8-11]. However, interfacial charge transport dynamics, especially across the metal-perovskite interface are not widely understood. Despite numerous reports related to devices based on metal-perovskite junctions, the transport dynamics have not been explored in detail [12-14]. Moreover, the study of carrier transport in the space charge regime is not conclusive. One of the reasons for this discrepancy is that the experimental studies are carried out on sandwich devices, where ionic motion affects carrier transport, even under equilibrium conditions. The present study demonstrates that lateral structure is a suitable framework to separate the effects of carrier transport and ionic motion. In this work, we study the carrier transport regimes in a HOIP based lateral metal-semiconductor-metal (MSM) device as a function of inter-electrode distance and applied bias. The lateral geometry provides access to spatially probe the transport parameters such as the local potential and photocurrent across the device. These spatially resolved measurements carried out in tandem with the bulk device characteristics provide a more comprehensive picture of the microscopic origins of carrier transport that affect the device response.

The transport studies are carried out on lateral back-contact MSM devices, with methylammonium lead iodide, $MAPbI_3$ (MAPI) as the hybrid-perovskite semiconductor. Unlike the widely used sandwich device architecture, lateral devices offer unique advantages such as: (i) Reduced light-absorption losses due to the absence of stack layers, (ii) low dark currents, which allow for small-signal detection, (iii) low device capacitance which reduces the RC lifetime, thereby enabling device operation for high-speed applications [15]. This work primarily investigates the transport characteristics of lateral asymmetric electrode devices with dissimilar work-function metals, i.e., Aluminum (Al) and Gold (Au). The choice of metal electrode work-functions, namely Al ($\varphi_m$ = 4.1 eV) and Au ($\varphi_m$ = 5.2 eV) in conjunction with MAPI allows for selective extraction of the photogenerated electron and hole, respectively. In the latter part of the work, we demonstrate the utility of lateral devices for photodetector application. Transient measurements on these devices show a high-speed response component of ~ 12 ns. These devices also exhibit a high linear dynamic range of ~ 118 dB, spanning close to six orders of light intensity variation.

## B. Results and Discussion

### 1. Device Operation

The lateral MSM (metal-semiconductor-metal) devices were fabricated using shadow masks, printed using a 2-photon-polymerization based 3D printer (Nanoscribe GmbH), and mask-liftoff techniques developed in the laboratory. After sequential deposition of Au and Al electrodes on the fused silica substrate, MAPI perovskite was deposited (~ 200 nm) to form the back contact lateral device, as shown in **Fig. 1(a)**. The absorption and PL spectra of MAPI films shown in **Fig. S1(a)** indicate a bandgap of ~ 1.6 eV. Additionally, the sharp XRD peaks (**Fig. S1(b)**) and the morphology (**Fig. S1(c)**) suggest high crystalline order in these films. A detailed explanation of device fabrication is provided in the experimental section of the Supplemental Material. As shown in the device schematic (**Fig. 1(a)**), $l$ is the inter-electrode channel length and $d$ is the thickness of the metal electrode (~ 80 – 100 nm), and $w$ is the width of the electrode. In this work, we study device characteristics as a function of inter-electrode channel lengths ($l$) in the range of 4 μm ≤ $l$ ≤ 120 μm. In these devices, the electrode width, $w \gg l$, is in the range of ~ 2 – 4 mm. The cross-sectional area of these devices is $w \times d$. **Fig. S2(a)** shows the reflection-microscopy image of the asymmetric electrode device, with $l$ ~ 12 μm. These devices can be implemented to a larger area using inter-digitated electrode patterns, as shown in **Fig. S2(b)**.

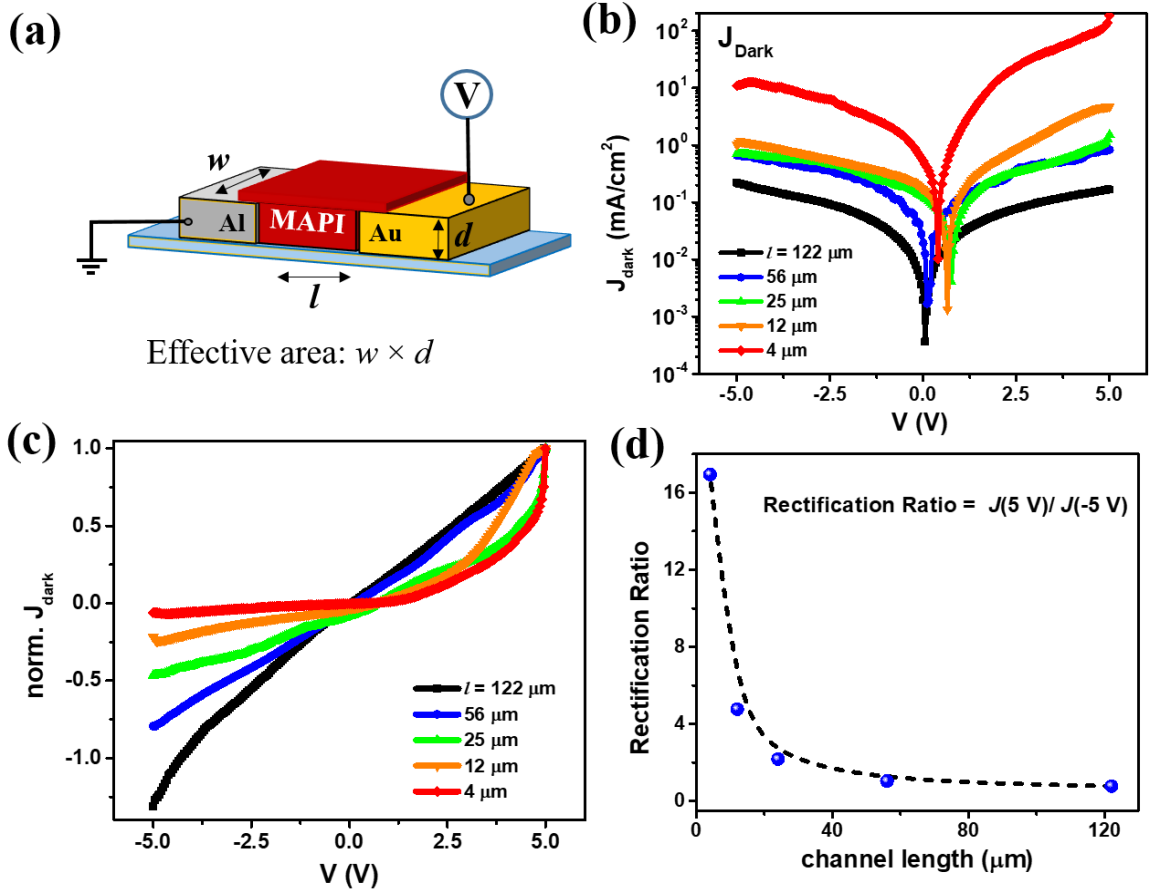

**FIG. 1**: **(a)** Schematic of the lateral asymmetric electrode device, where $l$ is the inter-electrode channel length and $d$ is the thickness of the metal electrode. **(b)** Semi-log plot of J-V characteristics of the asymmetric electrode device for different values of $l$. **(c)** The plot of the J-V characteristics, normalized to J at 5 V bias. **(d)** The plot of the rectification ratio as a function of channel length shows that the rectification ratio increases for short channel devices.

**Fig. 1(b)** shows the dark J-V characteristics on asymmetric MSM devices for different values of the inter-electrode distance, $l$. The magnitude of the dark-J at a given voltage increases for the short-channel devices owing to the higher electric field (V/$l$). Additionally, the J-V features indicate a rectification-like behavior in the regime of short-channel devices (**Fig. 1(c)**). Likewise, the rectification ratio = $\frac{J(5\,V)}{|J(-5\,V)|}$, shown in **Fig. 1(d)** shows a sharp increase for the short channel devices reaching up to ~ 17 for $l = 4$ μm. To understand the nature of the observed current, we carry out a closer analysis of the non-linear J-V characteristics.

2. **Carrier-transport regimes**

**Fig. 2(a)** and **2(b)** show the log-log plots of the J-V characteristics for devices with $l = 12$ μm and 4 μm respectively, in the positive bias sweep (to the Au electrode). The J-V features are linear (ohmic) at low voltages. Beyond a critical voltage, the current increases as $V^2$, suggestive of space charge limited current (SCLC) behavior and is described by [16,17]:

$$J_{SCLC} = \frac{9}{8}\mu\theta\varepsilon_0\varepsilon_r \frac{V^2}{l^3} \qquad (1)$$

Where $\mu\theta$ and $\varepsilon_r$ represent the SCLC mobility and the dielectric constant of the semiconductor, respectively. In the expression for the SCLC mobility: $\mu_{SCLC} = \mu\theta$, $\mu$ corresponds to the free carrier mobility and $\theta$ is the reduction factor due to the carrier-trapping effects [18]. The J-V characteristic is linear for longer channel length devices ($l > 12$ μm), as depicted in **Fig. S3**. However, the transition to the SCLC regime is expected to occur at much larger voltages in agreement with the scaling relation (**Eq. (1)**).

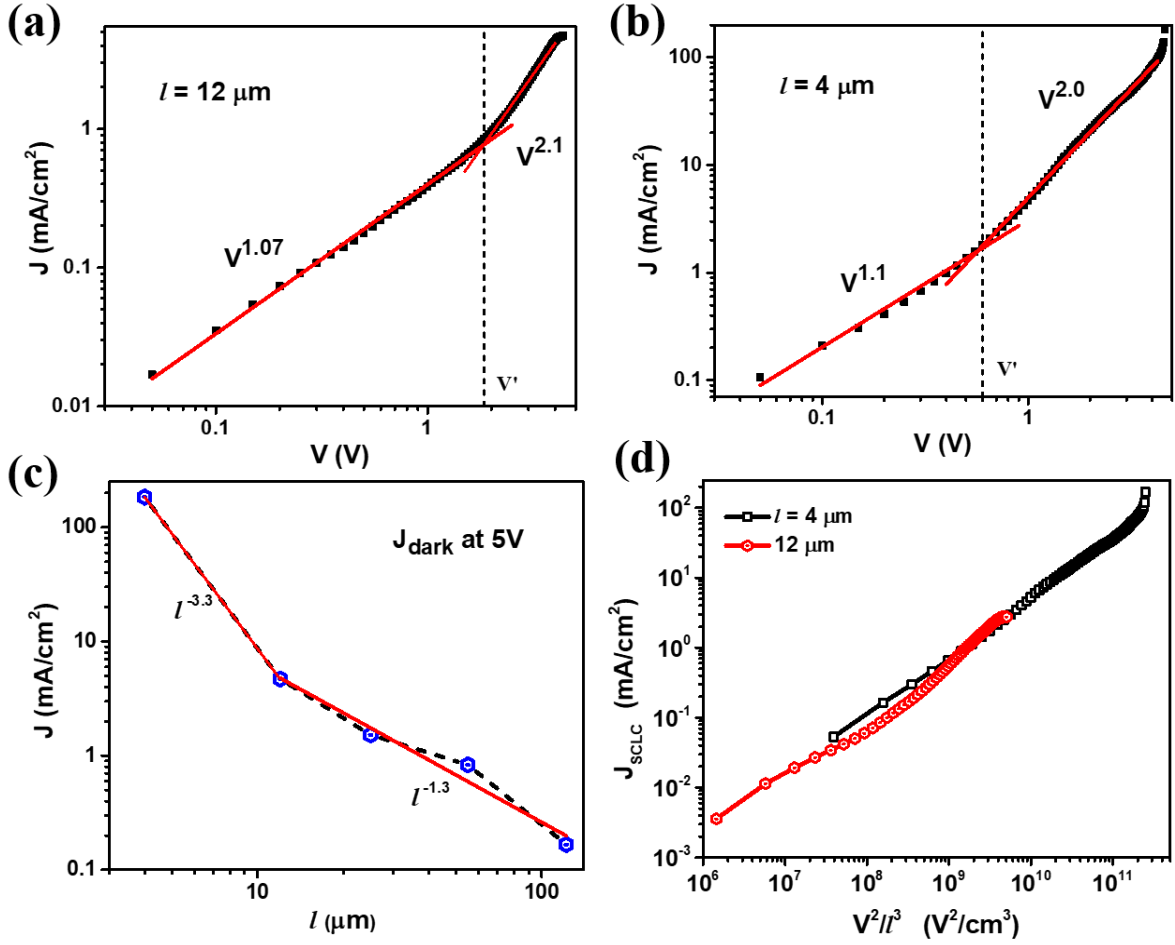

**FIG. 2**: Log-log plot of the dark J-V data shows distinct ohmic ($\propto V^1$) and SCLC behavior ($\propto V^2$) for **(a)** $l$ = 12 μm and **(b)** 4 μm channel respectively. **(c)** Variation of dark current density at 5 V as a function of channel length shows an ohmic behavior ($\propto l^{-1}$) for the long channel length and an SCLC behavior ($\propto l^{-3}$) for the short channel length. **(d)** Plot of current density, $J_{SCLC}$ as a function of $V^2/l^3$ for short channel devices shows a linear dependence, confirming SCLC behavior.

**Fig. 2(c)** represents the device current as a function of channel length at 5 V bias. The current is ohmic ($J \propto l^{-1}$) for long channel devices ($l > 24$ μm) and scales as $l^{-3}$ for short channel devices. If the transport is SCLC, $J$ should satisfy a scaling relation such that: $J_{SCLC} = f(V^2/l^3)$. The SCLC current, $J_{SCLC}$ was estimated considering the device current to be a combination of the ohmic and space charge limited current such that $J_{dark} = J_{ohmic} + J_{SCLC}$ [19] (Details related to the estimation of $J_{SCLC}$ are provided in section 12 of the Supplemental Material). **Fig. 2(d)** is a plot of $J_{SCLC}$ vs $V^2/l^3$, indicating a linear variation. The current density in the SCLC region is similar for both the $l = 4$ μm and 12 μm samples. This confirms that the transport in these samples is SCLC. Using **Eq. (1)** and $\varepsilon_r = 60$ [20,21], the SCLC mobility was determined to be in the range, $\mu_{SCLC} \approx 0.07 - 0.15$ cm$^2$/Vs. These values are consistent with previous observations of $\mu$ from FET and contact-based measurements [18,22-25]. The low lateral mobility in this system is attributed to scattering at the grain boundaries accompanied by carrier-trapping effects. Additionally, the plot in **Fig. S4** shows the linear J-V characteristics for positive potential sweep to Al electrode (negative bias sweep in **Fig. 1(b)**) for different values of the channel length. It is noted that this rectification-like characteristic is not observed in the case of symmetric electrode devices (**Fig. S5**). Therefore, the origin of this trend is attributed to the presence of a charge-selective transport barrier across either of the metal-perovskite interface. To investigate this further, we examine the energy levels and band-bending mechanism across the metal-perovskite interface.

### 3. Spatial Potential Mapping

**Fig. 3(a)** shows the band energy levels and the work-function of the metal electrodes in the absence of circuit formation. The work-function of Al and MAPI were determined to be 4.1 eV and 5.1 eV, respectively, using Kelvin-Probe measurements (with Au as the reference, details in the experimental section, Supplemental Material). The higher work-function of MAPI (5.1 eV) attributing a p-type character is consistent with previous observations [26,27], and the hole-carrier concentration is estimated to be ~ $10^{14}$ cm$^{-3}$. Upon forming a contact in short-circuit condition, the Fermi-level equilibrates, and band-bending occurs at the metal-perovskite interface, as shown in **Fig. 3(b)**. At the Al-MAPI interface, this results in an electron injection barrier $\varphi_{bn} = \varphi_m - \chi$ and a hole injection barrier, $\varphi_{bp} = I.E - \varphi_m$, where $\chi$ and $I.E$ is the electron affinity and ionization energy, respectively. Similar arguments govern the Au-MAPI interface, with an injection barrier for the electrons and holes represented as $\varphi'_{bn}$ and $\varphi'_{bp}$, respectively.

A detailed analysis for estimating barrier heights at the Al/MAPI interface and the Au/MAPI interface is elaborated in **Fig. S6**. This analysis indicates the presence of a finite injection barrier for both the carriers at the Al-MAPI interface. In contrast, the injection barrier for the holes at Au-HOIP is minimal (~ 0.1 eV) and can be assumed to be negligible. Additionally, a built-in potential is observed in the perovskite (in **Fig. 3(b)**), represented as $V_{bi}$ and $V'_{bi}$ at Al-MAPI and Au-MAPI interface, respectively, that arise as a result of band-bending at the interface. In the event of excess carrier generation, this built-in field facilitates electron and hole extraction to the Al and Au electrode, respectively.

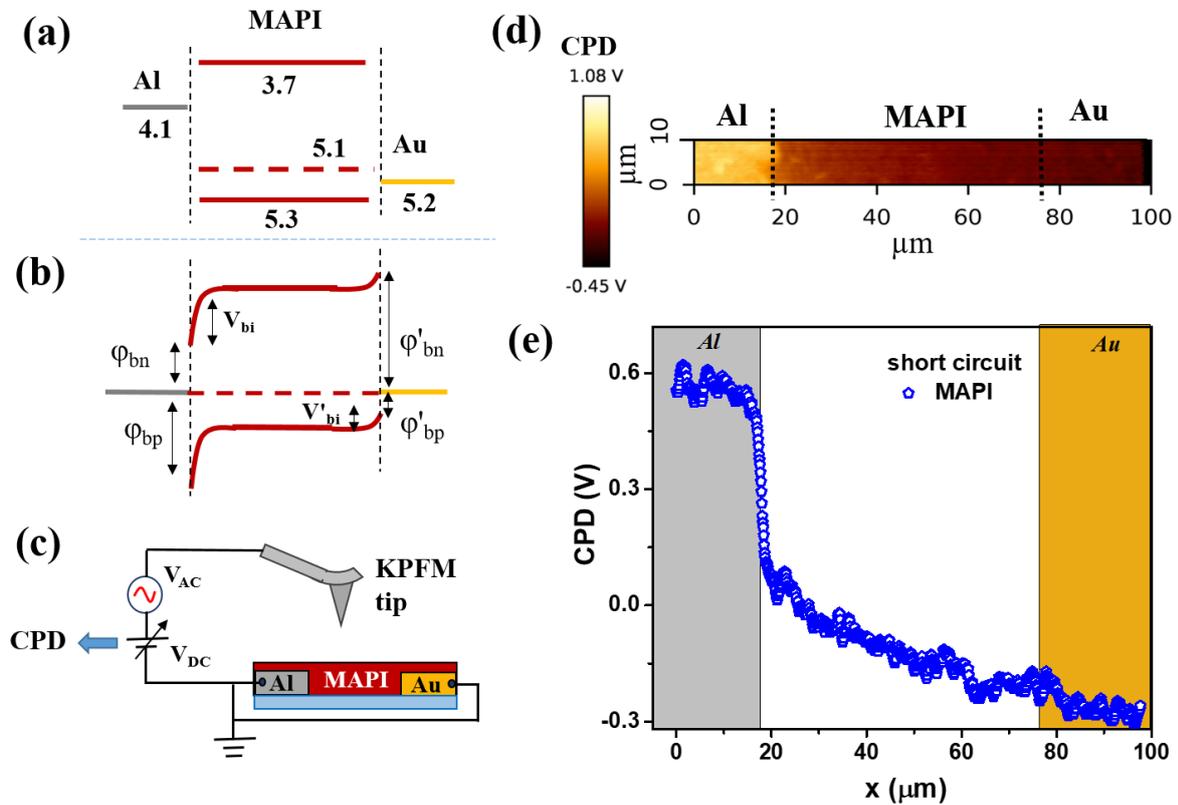

**FIG. 3: (a)** Band alignment diagram of the asymmetric electrode MSM device. **(b)** The device in short-circuit condition shows an injection barrier at the metal-perovskite interface. A built-in potential, $V_{bi}$ is developed at the metal-MAPI interface as a result of Fermi-level equilibration. **(c)** Schematic of the KPFM experimental setup used for potential mapping. The measured contact potential difference (CPD) is used to measure the potential profile across the device. **(d)** A 2D surface plot of the CPD measured across the device. **(e)** The CPD line-scan profile shows a significant potential drop at the Al-MAPI interface and across the bulk of the device. The potential drop across the Au-MAPI interface is minimal.

The lateral device structure offers the advantage to locally probe and map the potential using Kelvin Probe Force Microscopy (KPFM). The KPFM technique, depicted in **Fig. 3(c)**, essentially consists of a conductive tip that is in intermittent contact with the sample. As the tip is scanned across the device, a varied dc-potential with an overriding ac-bias is applied. At the point of Fermi-level equilibration between the tip and the sample, the contact potential difference (CPD), a parameter proportional to the surface potential is recorded. **Fig. 3(d)** shows the 2D surface plot of the CPD across the device (2D surface plots representing morphology and CPD correlation are presented in **Fig. S7**). In concurrence with the metal work-function, the CPD depicts a higher value for the Al electrode and a low value for the Au electrode. **Fig. 3(e)** shows the line scan profile of CPD across the device. The difference in the CPD value corresponding to the two electrodes is ~ 0.9 eV, which is in close agreement with the estimated value of 1.1 eV (**Fig. S6**). As expected, we observe a sharp potential drop across the Al-MAPI interface, which indicates the presence of a high $V_{bi}$ and an injection barrier at the interface. In contrast, at the Au-HOIP interface, the potential drop is minimal, suggesting a pseudo-ohmic contact for the hole carriers. Interestingly, across the bulk of the device, a gradual reduction of the CPD reveals the presence of a built-in electric field, indicative of negligible screening due to the mobile ions at the metal-perovskite interface. Any degrading process resulting from ion-accumulation at short-circuit operation is minimal [28]. Moreover, in the presence of low bias and low field conditions, degradation due to ionic drifts is not significant. It is to be noted that in our measurements on back contact lateral devices, the CPD corresponding to the metal surface is modified due to the presence of a thin layer (~ 50 nm) of the MAPI perovskite, as depicted in **Fig. 3(c)**. However, the spatial profile is indicative of the potential map across the lateral structure. This has been additionally verified by imaging the potential profile in the presence of an external bias.

The above-illustrated transport barrier at the Al-MAPI interface explains the observed rectification-like behavior in these asymmetric electrode devices. Essentially, in the negative bias condition (positive voltage to Al), an injection barrier is present for the holes and electrons at Al-MAPI and Au-MAPI interface, respectively. However, a high forward bias current is observed in the presence of a positive bias, with an efficient injection of holes across the Au-MAPI interface. Therefore the determined SCLC mobility corresponds to the injected hole-carriers. The asymmetry in the J-V (for short channel devices) shown in **Fig. 1(c)** then arises from the fact that Al does not make an ohmic contact with MAPI.

Previous reports on HOIP-based devices have shown that the classic SCLC formalism is invalid for systems with mobile ions [28-30]. The effect of ionic conduction is especially amplified in sandwich devices with an active layer thickness ~ 200- 300 nm. This results in a higher electric field (~ $10^5$ V/cm) with the dominant voltage drop across the metal-MAPI interface, thereby completely screening the bulk even in short-circuit condition [31,32]. On the other hand, in lateral structures, the low electric field (~ $10^2$ V/cm) across the bulk of the lateral device (**Fig. 3(e)**) suggests the minimal influence of ionic motion.

### 4. Spatial Photocurrent Mapping

We study the spatial photocurrent map using a narrowly localized excitation which is scanned across the lateral device. These results are correlated with the KPFM data to gain additional insight regarding photo-carrier transport in the background of the imaged potential landscape. Near-field Scanning Photocurrent Microscopy (NSPM) technique is employed to carry out these studies. The experimental setup, as depicted in **Fig. 4a**, consists of a tapered optical fiber-tip which is integrated with a tuning fork. Using a resonant-frequency-based feedback system, the tuning fork and the fiber-tip is lowered to be in close proximity to the device (experimental details provided in the Supplemental Material). With the sample-to-fiber tip distance in the near-field approximation (≤ $\lambda_{ex}$), beyond Abbe's diffraction limit, spot sizes up to ~ 100 nm can be achieved [33-35].

**Fig. 4(b)** shows the short-circuit photocurrent ($I_{ph}$) profile as the excitation is scanned across the asymmetric electrode device with $l$ = 22 μm. The $I_{ph}(x)$ profile indicates regions of low and high $I_{ph}$ for excitation close to the MAPI-Al and MAPI-Au interface, respectively. The higher $I_{ph}$ close to the Au electrode can be attributed to the efficient transport of the holes to the Au electrode. A model for understanding the reduction of the $I_{ph}$ close to the electrodes needs to be developed. It is to be noted that for local excitation, the magnitude of the $I_{ph}$ is also controlled by the dark resistance of the channel region outside of the illumination zone. The dark interfacial contact resistance is expected to be relatively higher in the short channel devices [36]. Possible origins of the low $I_{ph}$ close to the Al interface points to high recombination, presumably due to the presence of interfacial traps or a complex hetero-junction.

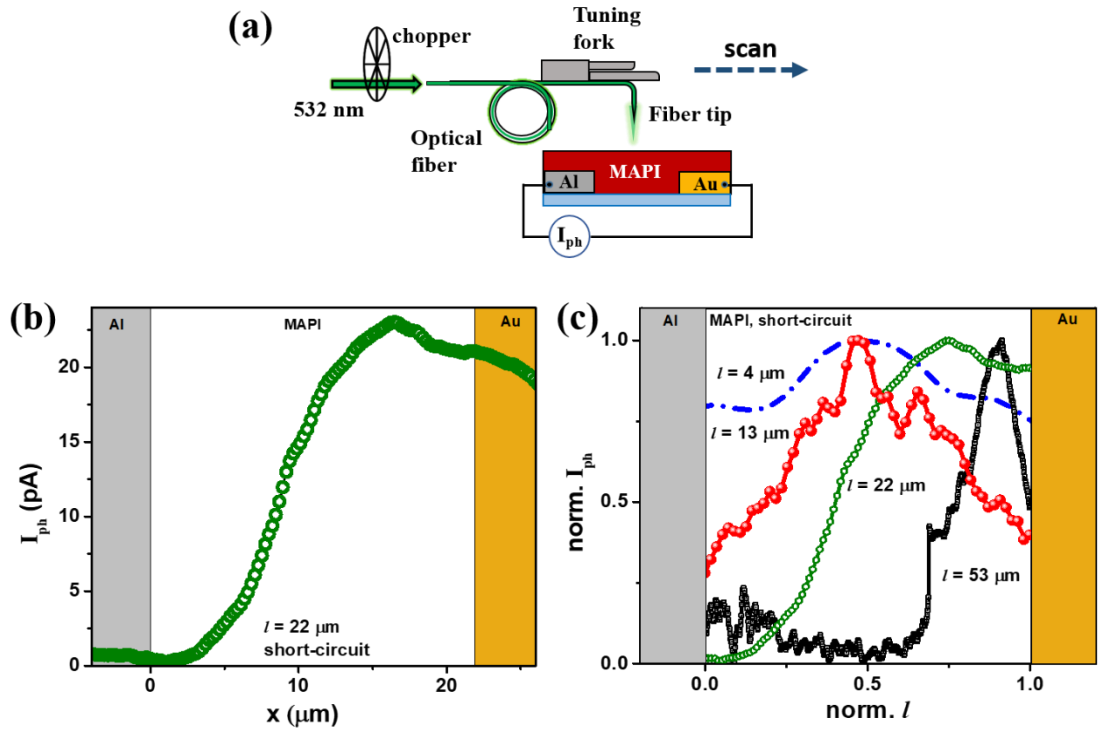

**FIG. 4**: **(a)** Schematic of Near-field Scanning Photocurrent Microscopy (NSPM) setup. A tapered fiber-tip in near-field excitation is scanned across the lateral device. **(b)** $I_{ph}(x)$ profile obtained for NSPM scan across $l$ = 22 μm MSM device shows the high $I_{ph}$ for excitation close to the Au/MAPI interface. **(c)**

Normalized $I_{ph}(x)$ profile as a function of channel length. For the short channel length devices, $I_{ph}(x)$ tends to a uniform distribution profile.

The influence on the $I_{ph}(x)$ profile was also studied as a function of $l$, results of which are presented in a normalized plot shown in **Fig. 4(c)** (raw data in **Fig. S8**). For longer $l = 56$ μm, the non-uniformity of the $I_{ph}$ distribution is skewed towards the MAPI-Au interface. However, for shorter channel $l = 13$ μm, we see that the $I_{ph}(x)$-peak is at the center, and the profile tends to a uniform spatial distribution. The dashed line for $l = 4$ μm, is the $I_{ph}(x)$ estimated using interpolation, also shows a higher degree of spatial uniformity.

### 5. Bias-dependence

To understand the microscopic transport features in the entire regime of device operation, we study the potential profiles in the presence of an applied bias. **Fig. 5(a)** shows the KPFM images of the lateral devices, with 5 V applied alternatively to the Al and Au electrodes (2D surface plots given in **Fig. S9**). In the presence of 5 V applied to the Al electrode, the measured net potential difference across the electrodes is > 5 V due to the added contribution of the built-in voltage, i.e., 5 V + $V_{bi}$. On the other hand, for the 5 V applied to the Au electrode, the net potential difference across the electrodes is < 5 V or 5 V - $V_{bi}$. Interestingly, the potential drop across the bulk of the device indicates a gradual reduction, suggesting a constant electric field (**Fig. S10**). In the case of negative bias (i.e., 5 V to Al), we observe a considerable potential drop across the MAPI-Al interface, indicative of a transport-barrier zone. The reduced dark-*J* in the reverse bias can be attributed to the inability of hole injection across this barrier. Previous works on lateral devices report dominant ion-migration effects under high electric fields over long time scales [37,38]. In our studies, experimental measurements were performed under conditions of low electric field (< $10^4$ V/cm) and shorter timescales to minimize the influence of ionic motion.

In the presence of an applied bias, the band-bending directionality allows for photo-carrier extraction to the respective electrodes. This photo-carrier extraction efficiency also depends on the carrier mobility and position of excitation from the collection electrode. **Fig. 5(b)** shows the $I_{ph}(x)$ profile obtained from the NSPM technique on the lateral devices in the presence of both positive and negative bias of 5 V. The profile indicates a spatial asymmetry, where the $I_{ph}$-peak shifts towards the positively biased (the electron extracting) electrode. This polarity-dependent $I_{ph}(x)$-asymmetry can be attributed to the unbalanced carrier mobility of the hole and the electron. To understand the microscopic picture of these bias-dependent photocurrent profiles, we simulate the carrier and photocurrent profiles using a simplistic 1D drift-diffusion formalism (details of simulation in section 13 of Supplemental Material). These results emphasize the dominant role of unbalanced mobility and can be obtained from solving the transport equation under conditions of steady-state generation at $x = 0$, drift condition assuming uniform electric field ($E = 1$ kV/cm) and hole mobility, $\mu_h = 100$ cm$^2$/Vs (this represents free carrier mobility in the ohmic regime of operation) [11]. **Fig. 5(c)** shows the simulated spatial distribution of the excess photogenerated electrons and holes. The spatial dependence of excess photogenerated carriers can be expressed as:

$$\delta n(x) = \delta n_0 \exp\left\{-\left(\frac{\sqrt{\left(\mu^2 E^2 + \frac{4D}{\tau}\right)} - \mu E}{2D}\right)x\right\} \quad (2)$$

where, $\delta n_0$, $E$, $D$ and $\tau$ represents the excess carrier density at the point of generation, electric field, diffusion coefficient and recombination lifetime, respectively. To account for unbalanced transport, we simulate carrier density profiles for different values of electron mobility, $\mu_e$, which is designated to be a fraction of the hole mobility. **Fig. 5c** shows that the electron distribution profiles exhibit a longer decay length for high $\mu_e$. Additionally, the directionality of hole and electron decay profiles show that holes drift in the direction of the electric field, and the electrons drift and decay opposite to the electric field direction. Further, to verify the profiles observed in the bias-dependent NSPM experiment (**Fig. 5(b)**), we simulate the spatial photocurrent density profile, $J_{ph}(x)$, which is given as a combination of diffusion and drift currents:

$$J_{ph}(x) = J_{drift} + J_{diffusion} = \delta n . e\mu E + eD\frac{\partial\, \delta n}{\partial x} \tag{3}$$

It is additionally noted that the observed current is a result of net $e - h$ recombination in the external circuit. **Fig. 5(d)** presents the simulated normalized photocurrent density profiles for various $\frac{\mu_e}{\mu_h}$ ratios. The profiles indicate that, in the case of balanced mobility, i.e., $\mu_e = \mu_h$, $I_{ph}$ peaks with excitation at the device center. This region of maxima denotes a zone for the optimal collection of electrons and holes at the respective electrodes. On the other hand, for unbalanced mobility ($\mu_e < \mu_h$), the peak shifts towards the electron-collection electrode. It is additionally noted that the carrier lifetime also affects the distribution profile (as indicated in **Eq. (2)**). Overall, the asymmetry of the $I_{ph}(x)$ points to the lower $\mu\tau$ product of the electrons ($\mu_e\tau_e < \mu_h\tau_h$). This interpretation is in agreement with earlier reports [11,39].

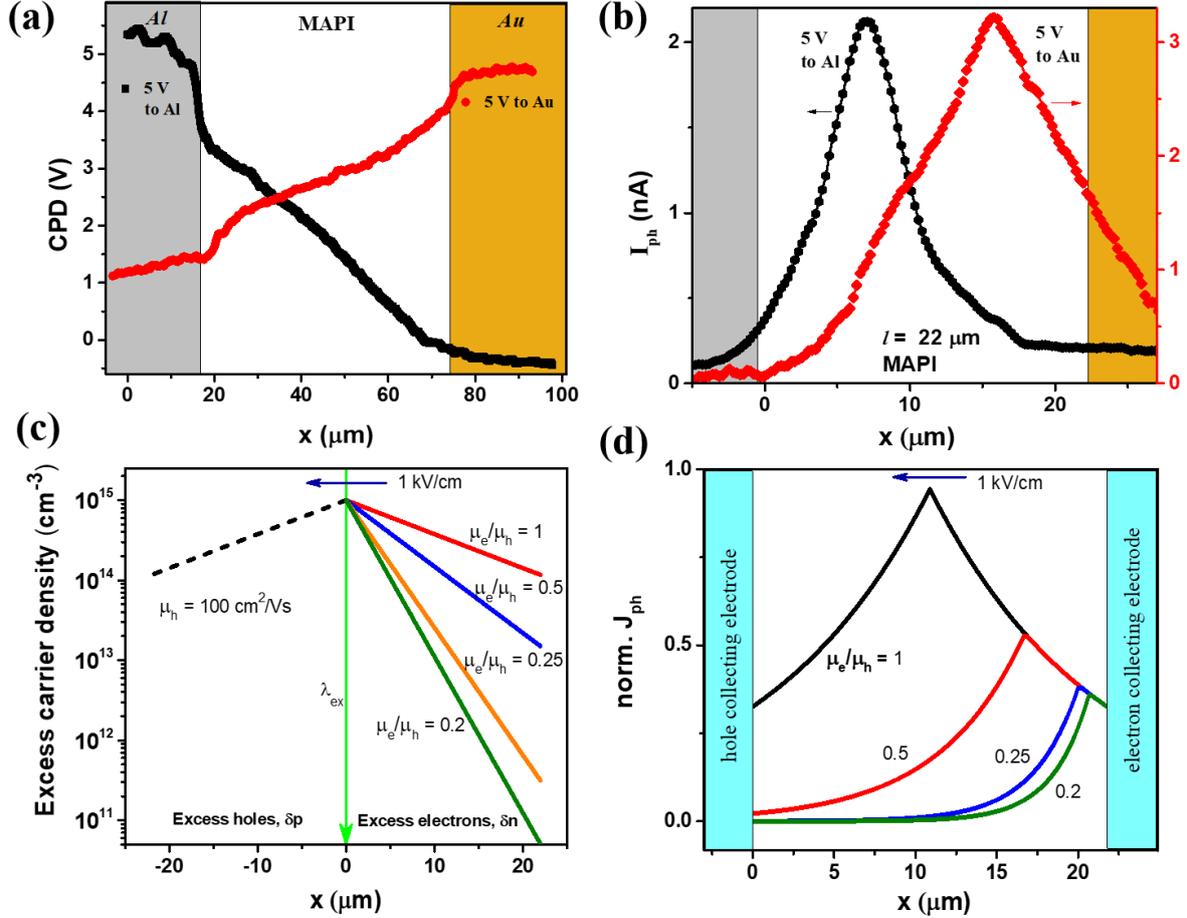

**FIG. 5**: **(a)** CPD profiles from KPFM on MSM devices for positive and negative bias of 5 V. Alternatively, this is indicated as 5 V applied to either Au or Al electrodes. The gradual potential drop indicates a uniform electric field across the device. **(b)** NSPM scans in the presence of external bias show the presence of a photocurrent peak which shifts depending on the polarity of the bias. This $I_{ph}(x)$ asymmetry indicates unbalanced carrier mobility. **(c)** Simulation using the finite element method shows the decay of excess electrons and holes for excitation at x = 0, under conditions of 1-D drift-diffusion transport for different $\mu_e$ values. **(d)** The normalized $J_{ph}$ profiles, simulated as a function of excitation position across the device, under 1 kV/cm electric field. The profiles indicate a shift in the $J_{ph}(x)$ peak in the case of unbalanced carrier mobility, which explains the feature observed in **(b)**.

### 6. Photodetector characteristics.

The light-dependent J-V characteristics on lateral MSM devices are shown in **Fig. 6(a)** under a 532 nm illumination condition (26 mW/cm$^2$). The $I_{ph}$ increases for the short-channel devices due to a higher electric field at any given voltage. It can also be seen that a short-circuit $I_{ph}$ is also observed corresponding to the values at V = 0. The directionality of the $I_{ph}$ confirms the electron extraction to the Al electrode and holes to the Au electrode. **Fig. 6(b)** shows the variation of the device responsivity with respect to the channel length. We achieved a responsivity, $R \approx 51$ mA/W in the self-powered mode and $R \approx 5.2$ A/W at 5 V operating voltage for the $l$ = 4 μm device. This corresponds to quantum efficiency, $QE = \frac{n_{carriers}}{n_{photons}} = \frac{R}{\lambda} \times 1240$ ($W.nm/A$) of 11.9 % and 1,212 % at 0 V and 5 V, respectively. The gain ($I_{ph}/I_{dark}$) observed in these devices is in the range of ~ $10^2$ (**Fig. S11**). The performance of these devices compares with lateral photodetectors reported in the literature [13,40,41].

The lateral devices are characterized by low device capacitance owing to the tiny overlap area across the two metal electrodes. This results in a small RC value, which reduces the overall response time of the device [42]. $I_{ph}(t)$ decay measurements on HOIP-based MSM devices have reported operational speeds in the range of ~ 40-90 ns [43]. We have additionally observed a fast decay component of ~ 12 ns (shown in **Fig. 6(c)**), using transient photocurrent measurements on MSM devices with mixed-phase perovskite: (FA$_{0.82}$MA$_{0.18}$Pb(I$_{0.82}$Br$_{0.18}$)$_3$) (FA = Formamidium) as the active semiconductor. Such fast-response times allow for applications involving high-speed detectors.

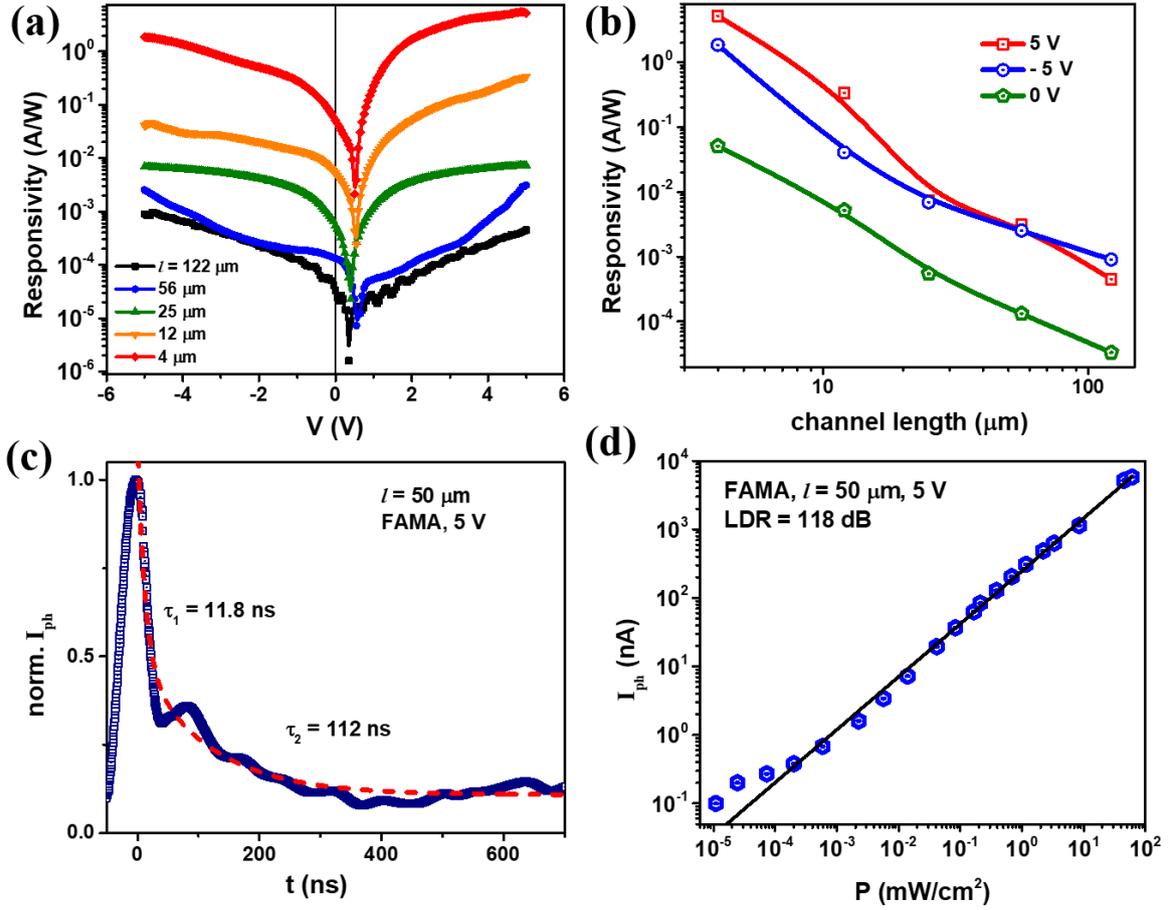

**FIG. 6**: **(a)** Voltage-dependent responsivity characteristics of the lateral devices for 532 nm illumination at 26 mW/cm$^2$. A short circuit $I_{ph}$ is observed at 0 V. **(b)** Variation of the responsivity as a function of channel length under short circuit conditions and ± 5 V bias conditions. **(c)** Transient photocurrent measurement on lateral MSM devices with mixed-phase perovskite (FAMA) as the active layer. The bi-exponential fit reveals a fast response time of 11.8 ns followed by a slow component of ~ 112 ns. **(d)** Variation of photocurrent with respect to light-intensity (λ = 532nm), shows the LDR ~ 118 dB, spanning over six orders of intensity variation.

The lateral device structure is characterized by a low dark current in comparison to the sandwich architecture. This reduces the noise level of device operation and allows for weak signal detection, increasing the dynamic range in these devices. **Fig. 6(d)** shows the linear dynamic range (LDR) of the lateral device with the mixed-phase perovskite in the active layer. The device maintains linear response over a range spanning ~ 6 orders of intensity variation, resulting in a $LDR = 20 \log \left(\frac{P_{max}}{P_{min}}\right)$ of 118 dB, where $P_{max}$ and $P_{min}$ correspond to the maximum and minimum light power, respectively, that is resolved with maintained responsivity.

### C. Conclusion

The lateral asymmetric electrode MSM devices exhibit distinct transport regimes as a function of applied bias and inter-electrode distance. The transport shows ohmic behavior at low voltages and becomes space charge limited at high voltage in the short channel length devices, consistent with the SCLC scaling relation. The observed rectification-like J-V characteristic is attributed to bias-direction dependent SCLC behavior due to efficient hole injection across the Au-MAPI interface. The KPFM plots across the device indicate an electric field in bulk, both in short-circuit and bias conditions, suggesting negligible ion-screening effects. In the presence of an applied bias, the spatial photocurrent profiles were understood in the context of drift-diffusion formalism, which indicates unbalanced carrier transport. The lateral devices exhibit photodetection capability and demonstrate a high responsivity of ≈ 5.2 A/W at 5 V operating bias in the short-channel structures. In addition to a higher dynamic range spanning six orders of intensity variation, the low device capacitance allows for high-speed applications in the GHz regime.

### D. Acknowledgments

The authors acknowledge the Department of Science and Technology, Government of India, and EPSRC-UKRI Global Challenge Research Fund project, SUNRISE (EP/P032591/1), for the financial assistance. SP acknowledges DST-INSPIRE for the fellowship.

**References**

[1] Q. Jiang, Y. Zhao, X. Zhang, X. Yang, Y. Chen, Z. Chu, Q. Ye, X. Li, Z. Yin, and J. You, Nature Photonics **13**, 460 (2019).

[2] N. R. E. Laboratory, (NREL Golden, CO, accessed June 2021).

[3] Y.-H. Lin, N. Sakai, P. Da, J. Wu, H. C. Sansom, A. J. Ramadan, S. Mahesh, J. Liu, R. D. Oliver, and J. Lim, Science **369**, 96 (2020).

[4] Y. Wang, Z. Lv, J. Chen, Z. Wang, Y. Zhou, L. Zhou, X. Chen, and S. T. Han, Advanced materials **30**, 1802883 (2018).

[5] Y. Wang, R. Fullon, M. Acerce, C. E. Petoukhoff, J. Yang, C. Chen, S. Du, S. K. Lai, S. P. Lau, and D. Voiry, Advanced Materials **29**, 1603995 (2017).

[6] C. Qin, A. S. Sandanayaka, C. Zhao, T. Matsushima, D. Zhang, T. Fujihara, and C. Adachi, Nature **585**, 53 (2020).

[7] L. Zhao, K. Roh, S. Kacmoli, K. Al Kurdi, S. Jhulki, S. Barlow, S. R. Marder, C. Gmachl, and B. P. Rand, Advanced Materials **32**, 2000752 (2020).

[8] Z. Chen, Q. Dong, Y. Liu, C. Bao, Y. Fang, Y. Lin, S. Tang, Q. Wang, X. Xiao, and Y. Bai, Nature communications **8**, 1 (2017).

[9] Y. Chen, H. Yi, X. Wu, R. Haroldson, Y. Gartstein, Y. Rodionov, K. Tikhonov, A. Zakhidov, X.-Y. Zhu, and V. Podzorov, Nature communications **7**, 1 (2016).

[10] S. D. Stranks, G. E. Eperon, G. Grancini, C. Menelaou, M. J. Alcocer, T. Leijtens, L. M. Herz, A. Petrozza, and H. J. Snaith, Science **342**, 341 (2013).

[11] Q. Dong, Y. Fang, Y. Shao, P. Mulligan, J. Qiu, L. Cao, and J. Huang, Science **347**, 967 (2015).

[12] S. J. Yang, M. Kim, H. Ko, D. H. Sin, J. H. Sung, J. Mun, J. Rho, M. H. Jo, and K. Cho, Advanced Functional Materials **28**, 1804067 (2018).

[13] J. Ding, S. Du, Y. Zhao, X. Zhang, Z. Zuo, H. Cui, X. Zhan, Y. Gu, and H. Sun, Journal of Materials Science **52**, 276 (2017).

[14] X. Lin, A. N. Jumabekov, N. N. Lal, A. R. Pascoe, D. E. Gómez, N. W. Duffy, A. S. Chesman, K. Sears, M. Fournier, and Y. Zhang, Nature communications **8**, 1 (2017).

[15] D. G. Georgiadou, J. Semple, A. A. Sagade, H. Forstén, P. Rantakari, Y.-H. Lin, F. Alkhalil, A. Seitkhan, K. Loganathan, and H. Faber, Nature Electronics **3**, 718 (2020).

[16] N. Mott, (Oxford University Press, 1940).

[17] M. A. Lampert and P. Mark, *Current injection in solids* (Academic press, 1970).

[18] M. S. Alvar, P. W. Blom, and G.-J. A. Wetzelaer, Nature communications **11**, 1 (2020).

[19] J. A. Röhr and R. C. MacKenzie, Journal of Applied Physics **128**, 165701 (2020).

[20] M. Zhang, X. Zhang, L.-Y. Huang, H.-Q. Lin, and G. Lu, Physical Review B **96**, 195203 (2017).

[21] N. Onoda-Yamamuro, T. Matsuo, and H. Suga, Journal of Physics and Chemistry of Solids **53**, 935 (1992).

[22] M. Petrovic, T. Ye, V. Chellappan, and S. Ramakrishna, ACS applied materials & interfaces **9**, 42769 (2017).

[23] D. Li, G. Wang, H.-C. Cheng, C.-Y. Chen, H. Wu, Y. Liu, Y. Huang, and X. Duan, Nature communications **7**, 1 (2016).

[24] X. Y. Chin, D. Cortecchia, J. Yin, A. Bruno, and C. Soci, Nature communications **6**, 1 (2015).

[25] S. P. Senanayak, B. Yang, T. H. Thomas, N. Giesbrecht, W. Huang, E. Gann, B. Nair, K. Goedel, S. Guha, and X. Moya, Science advances **3**, e1601935 (2017).

[26] J. R. Harwell, T. Baikie, I. Baikie, J. L. Payne, C. Ni, J. T. S. Irvine, G. A. Turnbull, and I. D. W. Samuel, Physical Chemistry Chemical Physics **18**, 19738 (2016).

[27] Y. Zhou and G. Long, The Journal of Physical Chemistry C **121**, 1455 (2017).

[28] E. A. Duijnstee, J. M. Ball, V. M. Le Corre, L. J. A. Koster, H. J. Snaith, and J. Lim, ACS Energy Letters **5**, 376 (2020).

[29] H. J. Snaith, A. Abate, J. M. Ball, G. E. Eperon, T. Leijtens, N. K. Noel, S. D. Stranks, J. T.-W. Wang, K. Wojciechowski, and W. Zhang, The journal of physical chemistry letters **5**, 1511 (2014).

[30] W. Tress, The journal of physical chemistry letters **8**, 3106 (2017).

[31] S. A. Weber, I. M. Hermes, S.-H. Turren-Cruz, C. Gort, V. W. Bergmann, L. Gilson, A. Hagfeldt, M. Graetzel, W. Tress, and R. Berger, Energy & Environmental Science **11**, 2404 (2018).

[32] H. Wang, A. Guerrero, A. Bou, A. M. Al-Mayouf, and J. Bisquert, Energy & Environmental Science **12**, 2054 (2019).

[33] C. R. McNeill, H. Frohne, J. L. Holdsworth, J. E. Furst, B. V. King, and P. C. Dastoor, Nano Letters **4**, 219 (2004).

[34] S. Mukhopadhyay, A. J. Das, and K. Narayan, The journal of physical chemistry letters **4**, 161 (2013).

[35] S. Mukhopadhyay, S. Ramachandra, and K. Narayan, The Journal of Physical Chemistry C **115**, 17184 (2011).

[36] M. Rao and K. Narayan, Applied Physics Letters **92**, 201 (2008).

[37] F. Liu, L. Wang, J. Wang, F. Wang, Y. Chen, S. Zhang, H. Sun, J. Liu, G. Wang, and Y. Hu, Advanced Functional Materials **31**, 2005662 (2021).

[38] Y. Yuan, J. Chae, Y. Shao, Q. Wang, Z. Xiao, A. Centrone, and J. Huang, Advanced Energy Materials **5**, 1500615 (2015).

[39] O. E. Semonin, G. A. Elbaz, D. B. Straus, T. D. Hull, D. W. Paley, A. M. Van der Zande, J. C. Hone, I. Kymissis, C. R. Kagan, and X. Roy, The journal of physical chemistry letters **7**, 3510 (2016).

[40] Y. Liu, Y. Zhang, K. Zhao, Z. Yang, J. Feng, X. Zhang, K. Wang, L. Meng, H. Ye, and M. Liu, Advanced Materials **30**, 1707314 (2018).

[41]　B. Yang, F. Zhang, J. Chen, S. Yang, X. Xia, T. Pullerits, W. Deng, and K. Han, Advanced Materials **29**, 1703758 (2017).

[42]　A. Armin, M. Hambsch, I. K. Kim, P. L. Burn, P. Meredith, and E. B. Namdas, Laser & Photonics Reviews **8**, 924 (2014).

[43]　F. X. Liang, J. Z. Wang, Z. X. Zhang, Y. Y. Wang, Y. Gao, and L. B. Luo, Advanced Optical Materials **5**, 1700654 (2017).

<u>**Supplemental Material**</u>

# Visualization of carrier transport in lateral metal-perovskite-metal structures and its influence on device operation

*N. Ganesh, A.Z. Ashar, Sumukh Purohit, K.L. Narasimhan and K.S. Narayan\**

## List of Contents:

# 1. MAPbI3 thin film characterization.

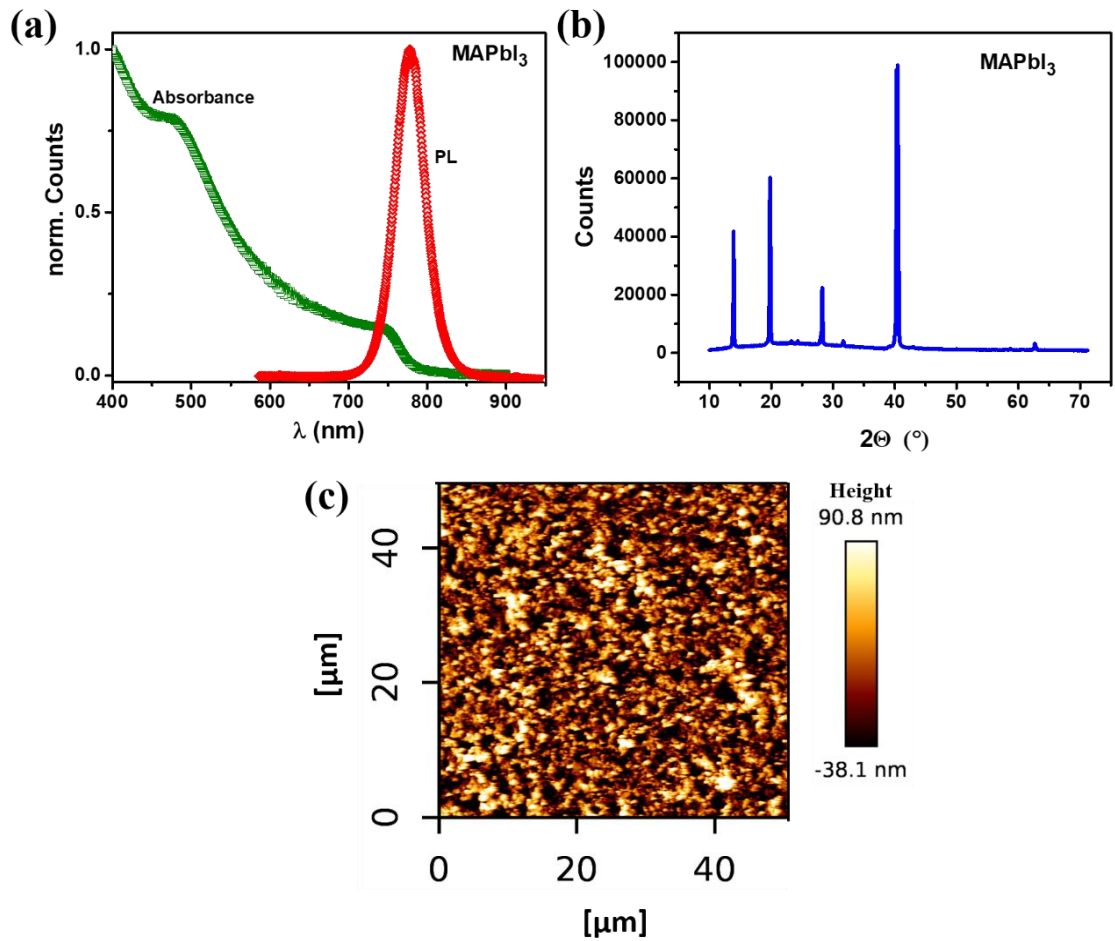

**FIG. S1**: **(a)** Absorption and PL data for the MAPbI$_3$ thin films. The absorption and Photoluminescence (PL) indicate a band-gap of 1.6 eV. **(b)** Data shows the thin film XRD of MAPbI$_3$ films. The sharp peaks indicate high crystalline order. **c.** Shows the morphology image of the MAPbI$_3$ films.

# 2. Microscopy image of asymmetric electrode device and inter-digitated device.

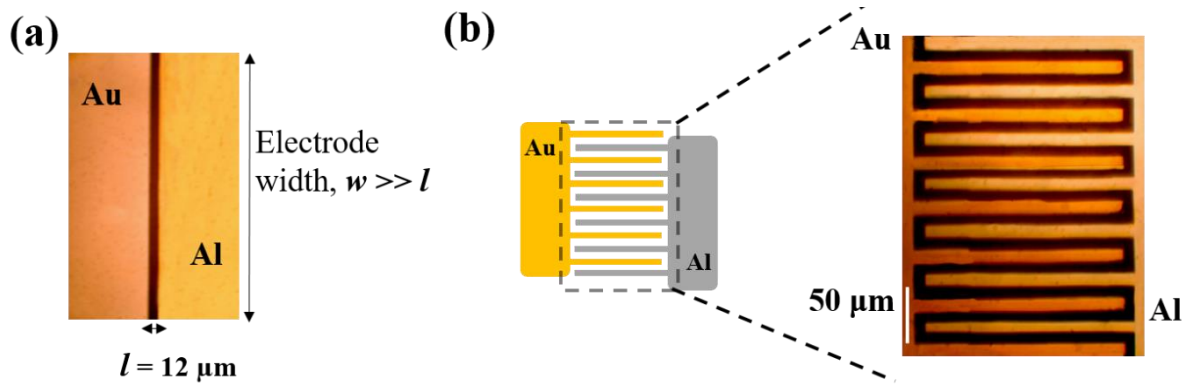

**FIG. S2**: **(a)** Microscopy image of the asymmetric electrode lateral device, with $l$ = 12 μm, captured in the reflection geometry. **(b)** Schematic of an asymmetric interdigitated electrode structure. The enlarged depiction is the microscopic image of the structure with an inter-electrode distance of 10 μm.

## 3. Log-log plot of positive bias J(V) characteristics for long channel devices.

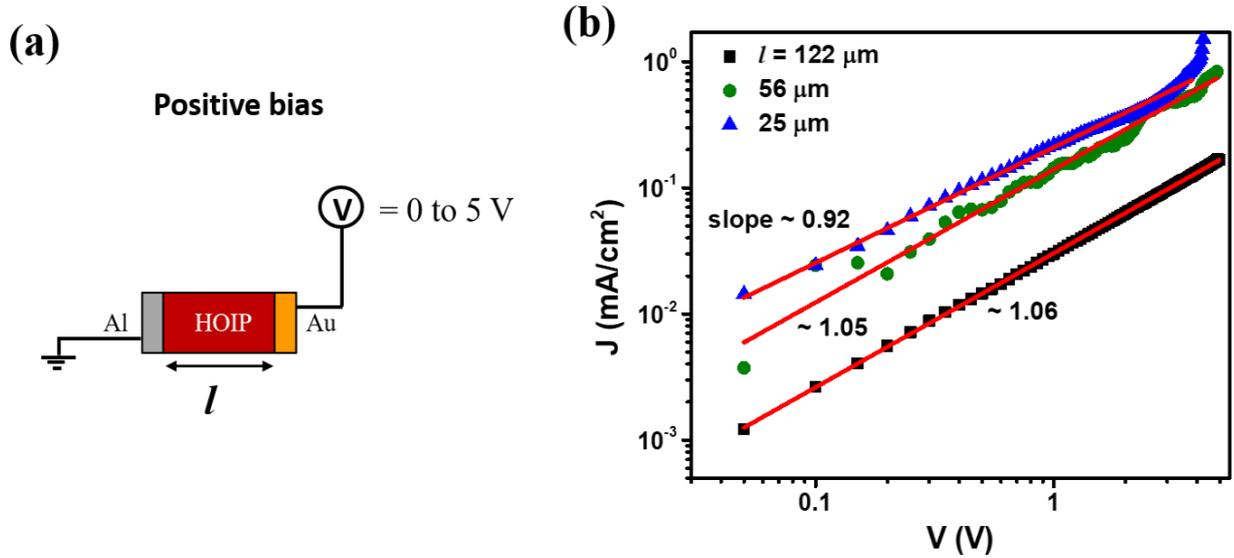

**FIG. S3**: **(a)** Schematic of the diagram representing the directionality of the applied bias. Here positive bias is applied to the Au electrode. **(b)** Log-log J(V) plot for long-channel ($l > 20$ μm) asymmetric electrode devices.

## 4. Log-log plot of negative bias J(V) characteristics of Al/MAPI/Au devices

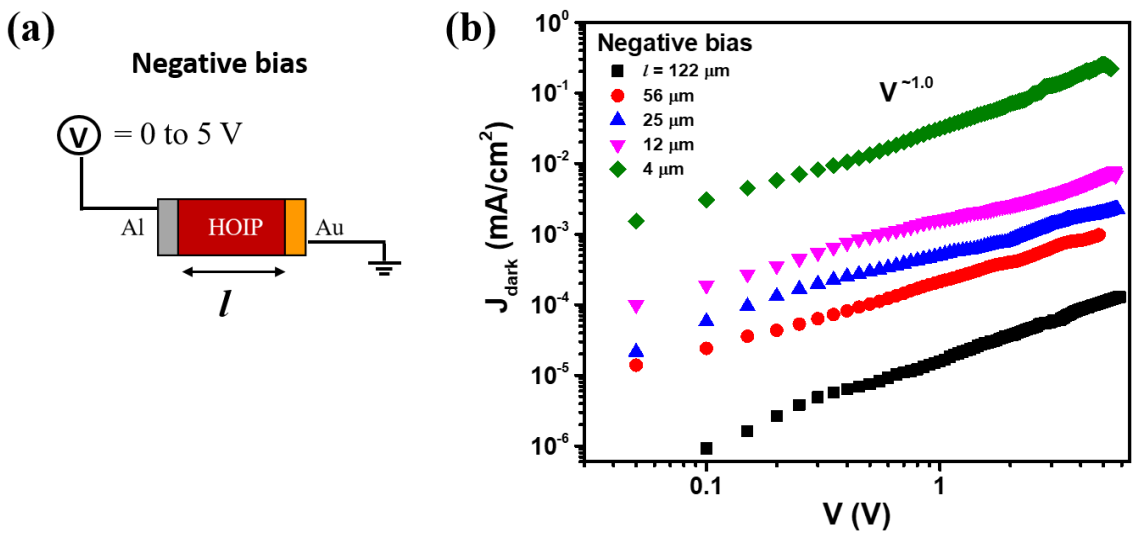

**FIG. S4**: **(a)** Schematic of the diagram representing the directionality of the applied bias. Here positive bias is applied to the Al electrode. **(b)** Log-log plot of dark J(V) characteristics show single-exponent variation, representing the ohmic regime of operation for devices of different channel lengths, $l$.

## 5. Dark J(V) characteristics of Au/MAPI/Au symmetric devices.

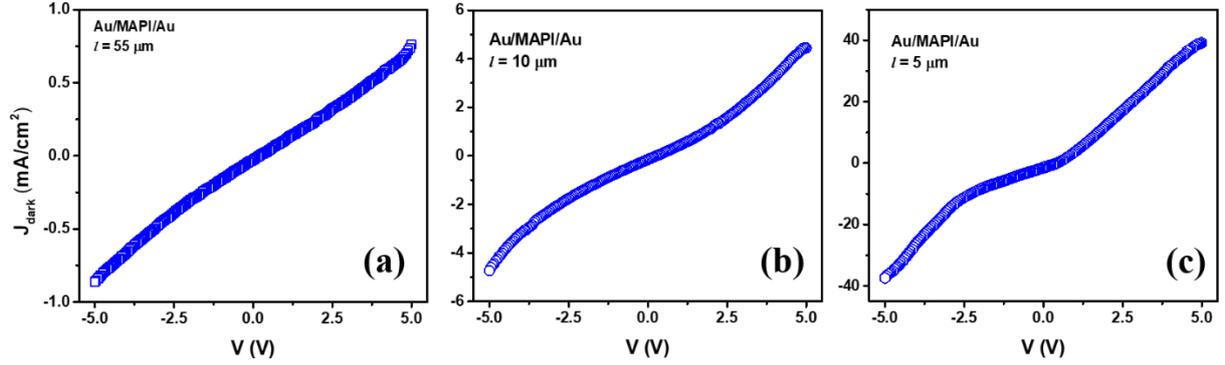

**FIG. S5**: Dark J(V) characteristics of Au/MAPI/Au symmetric electrode devices with a channel length of **(a)** 55 μm, **(b)** 10 μm, and **(c)** 5 μm.

## 6. Estimation of barrier height and $V_{bi}$ in asymmetric electrode device.

**Fig. S6** shows the band-alignment across the Al/MAPI (left-hand side of **Fig. S6**) and Au/MAPI (right-hand side of **Fig. S6**) interface. **Fig. S6(a)** and **S6(c)** show the Fermi-levels of the metals and the perovskite. The band-bending results in a barrier at the metal-perovskite interface.

At the <u>Al/MAPI</u> interface (**Fig. S6(b)**), the carrier injection barrier is given as:

$$\text{Electron barrier, } \varphi_{bn} = \varphi_m - \chi = 4.1 - 3.7 = 0.4 \text{ V}$$

$$\text{Hole barrier, } \varphi_{bp} = I.E - \varphi_m = 5.3 - 4.1 = 1.2 \text{ V}$$

Built-in voltage (favoring electron extraction): $V_{bi} = \varphi_{bn} - \varphi_n = 0.4 - (5.1 - 3.7) = -1.0$ V

At the <u>Au/MAPI</u> interface (**Fig. S6(d)**), the carrier injection barrier is given as:

$$\text{Electron barrier, } \varphi'_{bn} = \varphi_m - \chi = 5.2 - 3.7 = 1.5 \text{ V}$$

$$\text{Hole barrier, } \varphi'_{bp} = I.E - \varphi_m = 1.6 - 1.5 = 0.1 \text{ V}$$

Built-in voltage (favoring hole extraction): $V'_{bi} = \varphi'_{bn} - \varphi_n = 1.5 - (5.1 - 3.7) = 0.1$ V

The net built-in potential is given as: $V'_{bi}$ (Au/MAPI) – $V_{bi}$ (Al/MAPI) = **1.1 V**

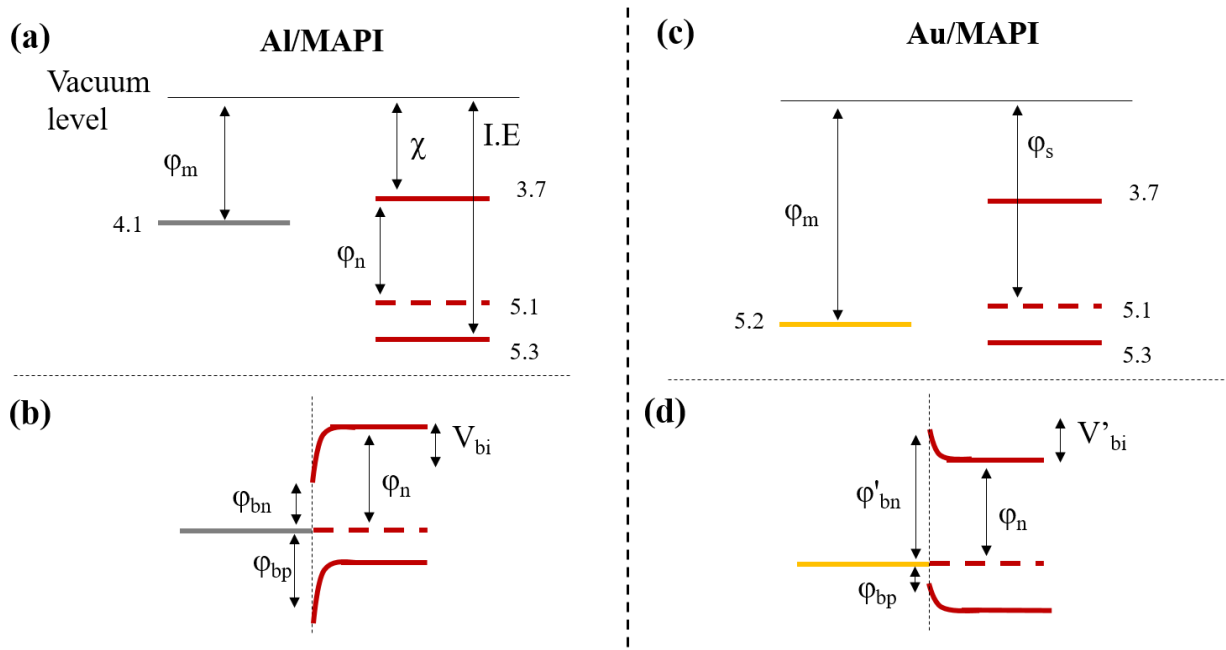

**FIG. S6**: Schematic shows the band alignment diagram across the Al/MAPI interface **(a)** before and **(b)** after the formation of the contact. The right-hand side shows the band-diagram across the Au/MAPI interface **(c)** before and **(d)** after the formation of the contact.

7. **Morphology-KPFM correlation of asymmetric electrode devices.**

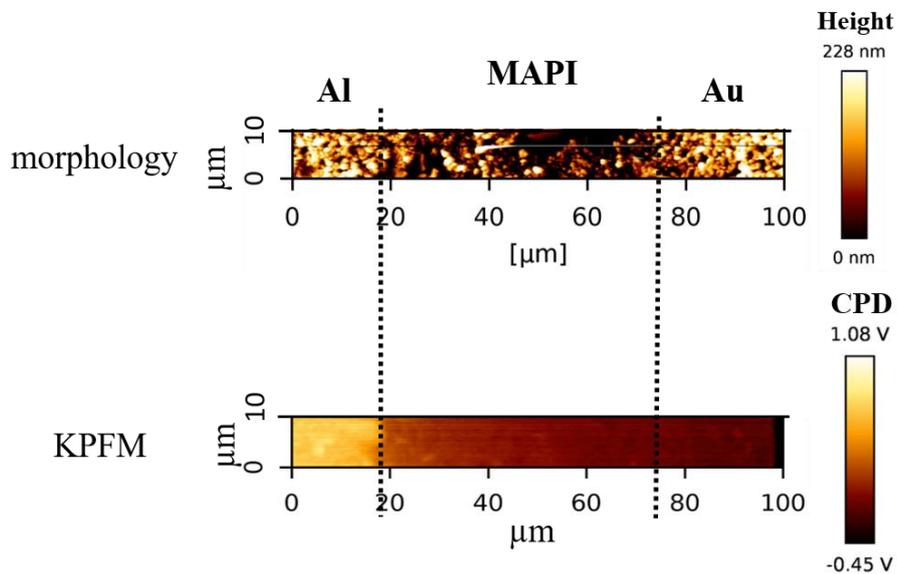

**FIG. S7**: 2D plots showing the morphology-KPFM correlation for asymmetric electrode device with $l$ = 55 μm.

## 8. Line profiles of NSPM scan in asymmetric devices for different *l*.

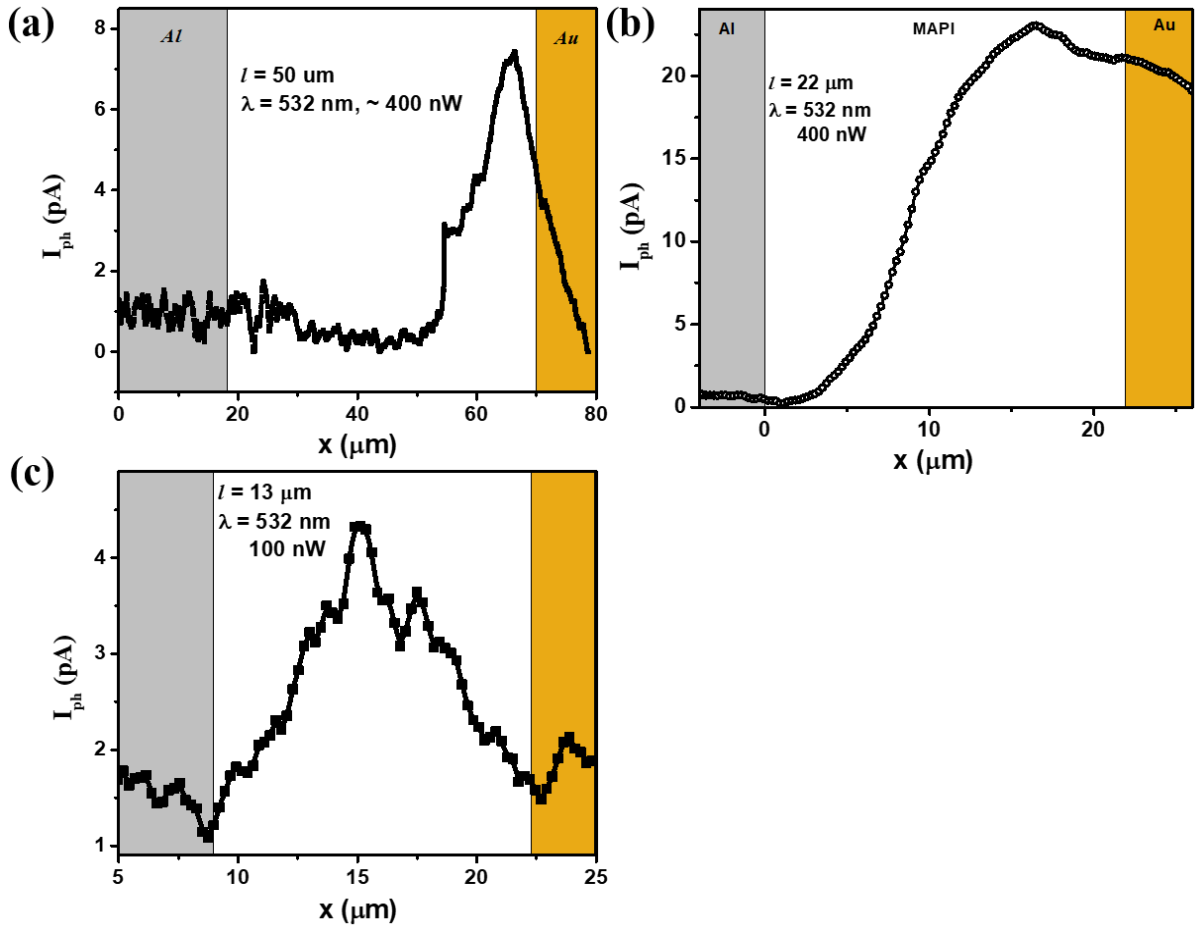

**FIG. S8**: Line scan profiles of the NSPM scans on asymmetric electrodes in short-circuit condition for the inter-electrode channel length of **(a)** 55 μm **(b)** 22 μm and **(c)** 13 μm.

## 9. 2D surface plots for bias-dependent KPFM

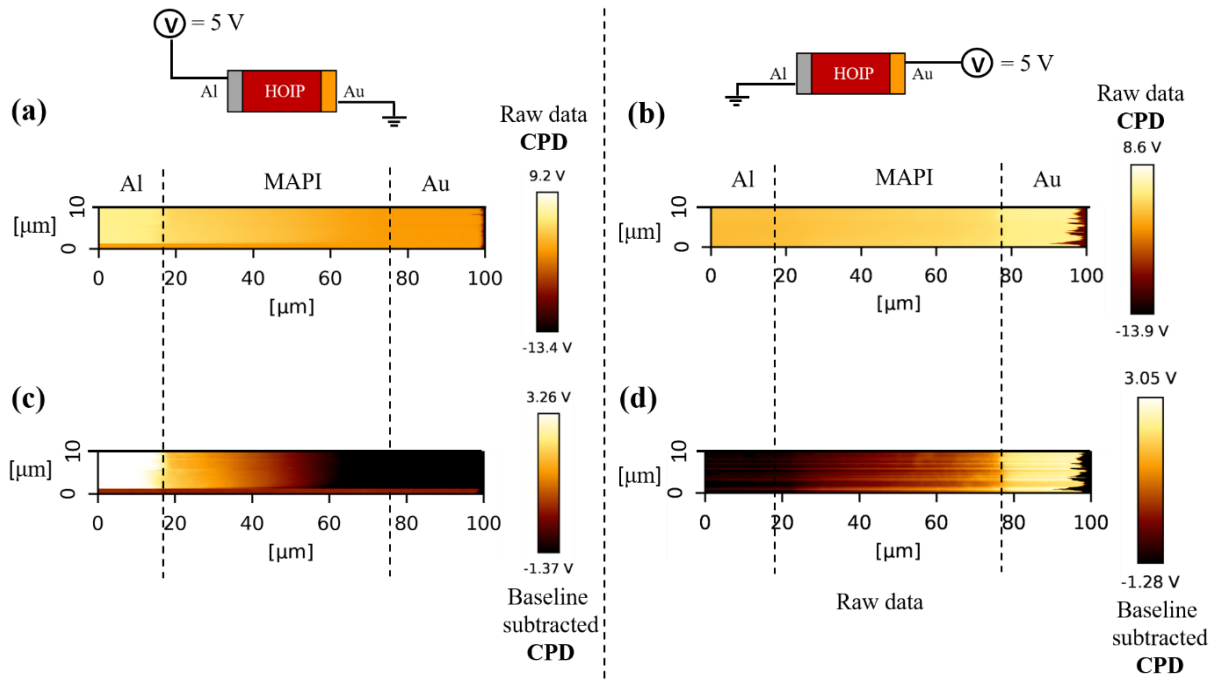

**FIG. S9**: 2D surface plots showing the raw data of the CPD for bias-dependent KPFM measurement with (**a**) 5 V applied to the Al and (**b**) 5V applied to Au electrode. (**c**) shows the data for 5 V applied to Al with additional data processing by subtraction of a zeroth-order polynomial baseline to improve the contrast of the image. (**d**) shows the data for 5 V to Au electrode, with similar data processing. However, only the raw data was used for all the analysis presented in the main text.

## 10. Electric-filed profile in the presence of bias.

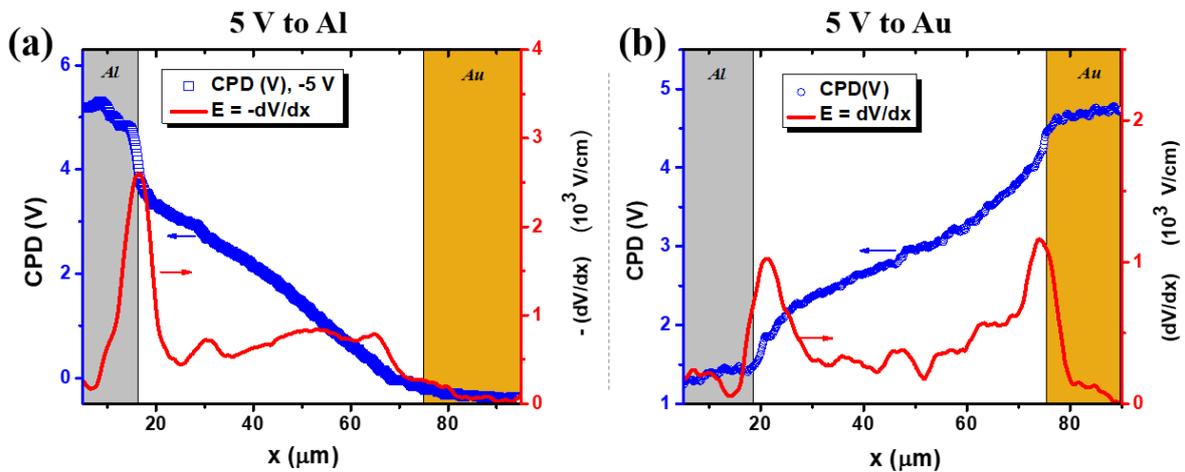

**FIG. S10**: Plot of CPD and lateral electric field for asymmetric electrode devices with $l = 56$ μm for the case of (**a**) -5 V and (**b**) + 5 V. The dominant electric field at the Al-MAPI interface in the negative bias is indicative of the formation of a depletion zone.

## 11. Comparison of dark and light I-V characteristics

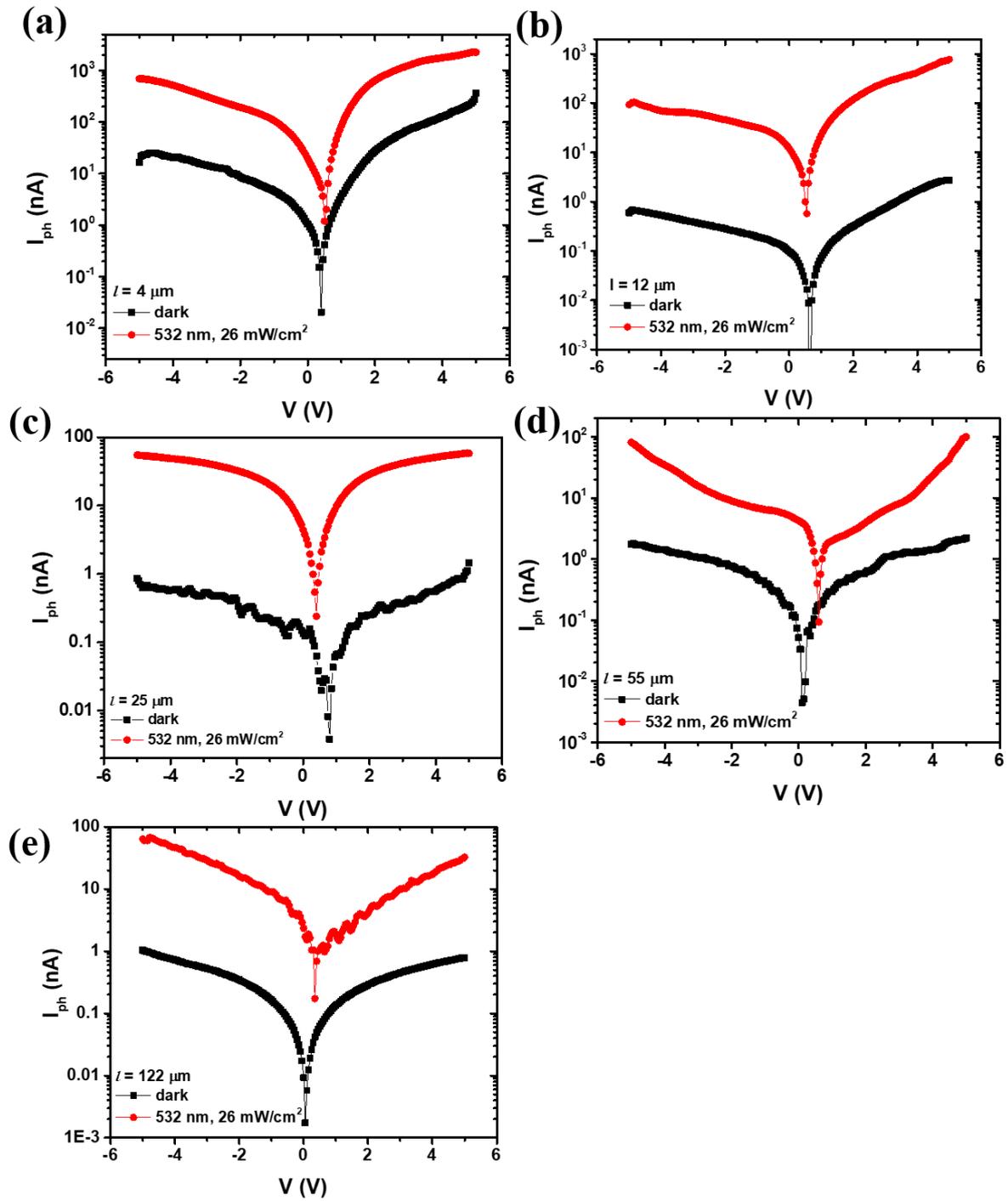

**FIG. S11**: Comparison of the dark and light (532 nm, 26 mW/cm$^2$) for channel lengths of **(a)** 4 μm, **(b)** 12 μm, **(c)** 25 μm, **(d)** 55 μm and **(e)** 122 μm

## 12. Representation of the SCLC current correcting for the ohmic contribution.

**Fig. 2(d)** (in the text) shows the plot of $J_{SCLC}$ vs. $V^2/l^3$. $J_{SCLC}$ corresponds to the SCLC dark current corrected for the ohmic contribution. For instance, **Fig. S12** shows the log-log J(V) plot for a 4 μm channel device with positive bias. The profile indicates the dominant ohmic and SCLC regimes as a function of the function of applied bias. Here, the ohmic current is given as $J_{ohmic} = n_0 e \mu \frac{V}{L}$, and $J_{SCLC} = \frac{9}{8} \mu \theta \varepsilon_0 \varepsilon_r \frac{V^2}{L^3}$ is given by the Mott-Gurney Law [1, 2]. The total mixed current can be represented as [3]:

$$J = J_{ohmic} + J_{sclc} \quad (S1)$$

To account for only the SCLC part of the observed current, at bias beyond the SCLC onset voltage, V', the SCLC current is given as:

$$J_{sclc} = J - J_{ohmic} \quad (S2)$$

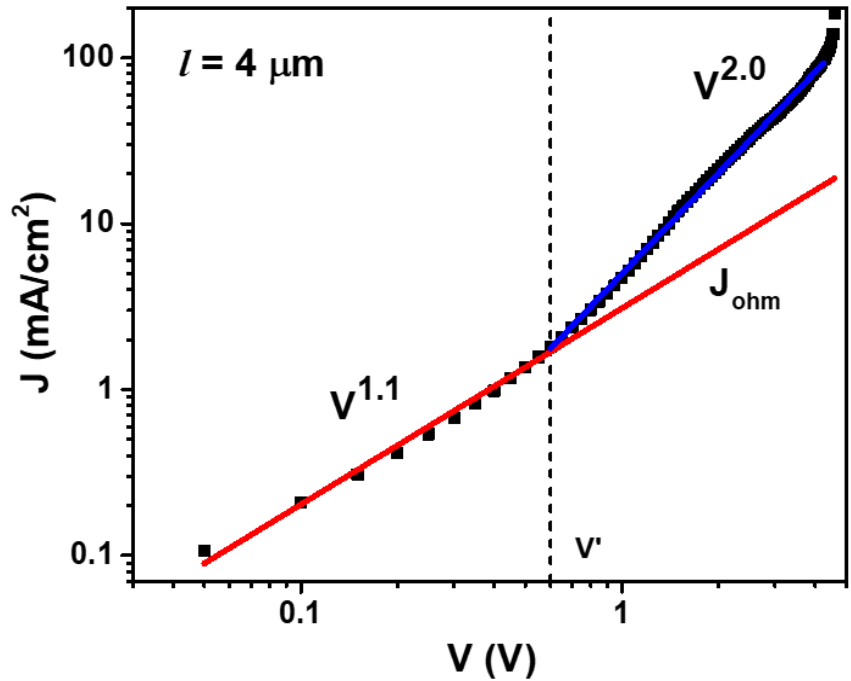

**FIG. S12**: Log-log plot of the J(V) characteristics for positive bias sweep in a $l$ = 4 μm asymmetric device. The plot shows distinct ohmic and SCLC behavior. The solid red line shows the baseline for the ohmic current.

From the plot in **Fig. S12**, the extrapolated red line corresponds to the $J_{ohmic}$, which is corrected, using **Eq. (S2)**, to obtain the SCLC current, $J_{SCLC}$. It must be noted that the voltage represented in **Fig. 2(d)** corresponds only to the SCLC voltage, i.e., V > V'.

### 13. Modeling and simulation of carrier drift in MSM devices.

To model the carrier drift in the lateral devices, we solve the transport equation. The transport of excess electrons is given as:

$$\frac{\partial \delta n}{\partial t} = D \frac{\partial^2 \delta n}{\partial x^2} + \mu E \frac{\partial \delta n}{\partial x} + G - \frac{\delta n}{\tau} \qquad (S3)$$

where, $\delta n_0$, $E$, $D$, $G$ and $\tau$ represents the excess carrier density at the point of generation, electric field, diffusion coefficient, generation function, and recombination lifetime, respectively. In our bias-dependent experiments, we have an electric field in the range of $E \sim 1$ kV/cm. In the scanning photocurrent experiments, where we have a steady-state point generation of photo-carriers, the carrier generation function is given $G(x) = G_0 \delta(x-x_0)$. Solving for 1D transport with carrier collection away from the point of generation ($x \neq x_0$), the **Eq. (S3)** reduces to the following second-order homogeneous equation:

$$D \frac{\partial^2 \delta n}{\partial x^2} + \mu E \frac{\partial \delta n}{\partial x} - \frac{\delta n}{\tau} = 0 \qquad (S4)$$

We solve for the excess carrier concentration by implementing the initial value conditions such that such that $\delta n(x = 0) = \delta n_0 = aG_0\tau$, and $\delta n(x = \infty) = 0$. Additionally, it is noted that the excess electrons drift in the direction opposite to the electric field, $-E$. At the charge-selective contacts, the boundary conditions for the excess electrons ($\delta n$) and holes ($\delta p$) can be assumed as follows:

(i) The Al electrode act as electron extraction contacts and reflects the holes such that:

$$\delta n(x = Al) = \delta n(x)$$

$$\delta p(x = Al) = 0$$

(ii) The Au electrode extracts holes and reflects electrons

$$\delta p(x = Au) = \delta p(x)$$

$$\delta n(x = Au) = 0$$

The excess carrier concentration profile, obtained as a solution to **Eq (S4)**, is given as:

$$\delta n(x) = \delta n_0 \exp\left\{-\left(\frac{\sqrt{\left(\mu^2 E^2 + \frac{4D}{\tau}\right)} - \mu E}{2D}\right) x\right\} \quad (S5)$$

The spatial profiles using the above expression for excess electrons and holes are given in **Fig. 5(c)**. Here we have considered the hole mobility as $\mu_h$ = 100 cm$^2$/Vs, recombination lifetime $\tau$ = 10 ns, an electric field, $E$ = 1 kV/cm, and $\delta n_0$ = 10$^{15}$/cm$^3$. The electron distribution, shown in **Fig. 5(c)** is simulated for different values of electron mobility, $\mu_e$, which is designated as a fraction of $\mu_h$, in accordance with previous reports [4, 5].

To simulate the NSPM experiment, the photocurrent due to carrier drift and diffusion is given as:

$$J_{ph}(x) = J_{drift} + J_{diffusion} = \delta n. e\mu E + eD\frac{\partial \delta n}{\partial x} \quad (S6)$$

The net current density is considered to be equivalent to the excess electrons and holes that recombine in the external circuit. This depends on the distribution profile of both the carriers. It can be seen in **Fig. 5(d)** that for the case of balanced mobility ($\mu_h = \mu_e$) the simulated photocurrent density profile exhibits a peak for photo-excitation at the device center, indicating an efficient collection of both the electrons as well as the holes. In the case of unbalanced carrier mobility ($\mu_h > \mu_e$), the I$_{ph}$ peak shifts towards the electron-extracting electrode.

## 14. Experimental Section

*Fabrication of asymmetric metal electrodes:*

The lateral metal-electrodes were deposited using 3D mask-printing (*Nanoscribe GmbH*). The masks were printed with a 2-photon-polymerization lithography technique using photo-curable resin (IP Dip) drop cast on fused silica substrates. After the development of the photoresist mask, Au metal was thermally evaporated. The resist was then removed by immersing in liquid $N_2$ for 30 s and then blow dry with air. The second mask corresponding to the channel length (and partially masking the edge of the Au electrode) was then printed using optical markers and then developed. This was followed by thermal evaporation of the Al electrode. Finally, the second mask was also removed to form the asymmetric electrode device. The thickness of the metal coating was in the range of 80 -100 nm.

*Fabrication of lateral back contact devices.*

The $MAPbI_3$ perovskite films were prepared by spin-coating the precursor solution. The $MAPbI_3$ perovskite precursor was initially prepared using the following by dissolving 357 mg of methylammonium iodide (MAI, Dyesol), 204 mg lead chloride ($PbCl_2$, Sigma Aldrich) and 7 mg of lead iodide ($PbI_2$, 99.999 %, Sigma Aldrich) in 1 ml of DMF (N,N-Dimethylformamide, anhydrous) solution and stirred at high speed for at least 60 min. To render the substrates hydrophilic, they were plasma treated (using ultra-high pure (UHP) nitrogen, 99.9995 %) for 2 min. The substrates were transferred to a glovebox maintained in an inert atmosphere. The perovskite precursor was dispensed onto the substrates and spin-coated at 3000 rpm for 80 s followed by 5000 rpm for 10 s. The films were then annealed in the following sequence: 20 °C for 15 min, 70 °C for 15 min, 100 °C for 90 min and 120 °C for 10 min. To encapsulate the devices (for J-V measurements), a layer of PMMA (250 mg ml$^{-1}$ in chlorobenzene) was spin-coated at 2000 rpm for 60 s followed by annealing at 60 °C for 20 min.

For the preparation of mixed-phase perovskite precursor solution, 215 mg of FAI (Formamidium Iodide, Sigma Aldrich, > 99 %), 28 mg of MABr (Methylammonium bromide,

Dyesol), 634 mg of PbI$_2$ (Lead(II) iodide, Alfa Aesar, 99.9985 %) and 92 mg of PbBr$_2$ (Lead Bromide, Sigma Aldrich, 98 %) was dissolved in 1 ml of 4:1 ratio of (Dimethylformamide : Dimethylsulfoxide) and stirred at 85 °C for about 30 min. 70 μL of the solution was then dispensed onto the substrates with lateral metal electrodes and spin-coated at 1000 rpm for 10 s followed by 6000 rpm for 35 s. In the last 7 s of the spin, 70 μL of Anisole (anhydrous, Sigma Aldrich) was dynamically dispensed. The films were then annealed at 100 °C for 60 min. All the materials were used as obtained.

*J-V measurement on devices*

The J-V measurements were carried out on devices using the Keithley Source Meter 2400. The scan on the voltage sweep was carried out at a rate of 0.5 V/s. The light responsivity measurements were carried out using a 532 nm excitation with an optical power of ~ 26 mW/cm$^2$.

Several devices (> 20) were fabricated for the studies. The dark I-V studies presented were analyzed from three devices each corresponding to $l$ = 25 μm, 50 μm and 100 μm and from two devices corresponding to channel length, $l$ = 4 μm and 12 μm, and. These devices represented different batches during the fabrication process. We consistently observe that the long channel devices show symmetric I-V characteristics while the short channel devices exhibit rectification-like behavior. These devices were also utilized for the photo-detector studies.

*Kelvin Probe Force Microscopy*

Kelvin Probe measurements were carried out using the JPK Nanowizard-3 Atomic Force Microscopy instrument. The scans were performed using conductive Cr-Pt cantilever tips (BudgetSensors, Multi 75 EG) with a force calibration of 2 N/m. Calibration of the tip was initially done by measuring the contact potential difference (CPD) on the Au sample (100 nm of Au coated on the glass slide). This was followed by a measurement of the contact potential difference on the lateral devices. The thickness of the perovskite layer was reduced to ~ 150-180 nm to minimize masking effects due to the thick layer and capture the potential profile of the active device. In short-circuit condition, both the Al and Au electrodes were grounded with respect to the Kelvin-Probe circuit. In the case of bias-dependent studies, an external bias was applied in parallel to the lateral MSM device. For the spatially resolved studies, we fabricated and measured devices with channel lengths of $l$ = 22 μm and 55 μm. Here a minimum of two devices of each was used, and the measurements were carried out at numerous points across the channel length.

*Near-field Scanning Photocurrent Microscopy*

JPK Nanowizard-3 Atomic force microscope (AFM) in Near-field scanning optical microscopy (NSOM) mode was used to measure the local photocurrent in the channel region of the devices. A tapered optical fiber coated with Cr/Au and having aperture ~ 105 nm was raster-scanned using the piezo head of the AFM. The other end of the multi-mode fiber was coupled to a 532 nm laser using a 20x, N.A.= 0.4 microscope objective lens. The input laser power was maintained at 5 mW during the measurement. Additionally, the laser intensity was modulated at 383 Hz using a mechanical chopper. Using a resonant-frequency based feedback system, the assembly consisting of the tuning fork and the fiber-tip was lowered to be in close proximity to the sample. When the fiber tip-to-sample distance is in the near-field approximation ($\leq \lambda_{ex}$), spot sizes up to ~ 100 nm can be achieved beyond Abbe's diffraction limit.

In synchronization with the excitation scanning, the photocurrent signal was measured using Stanford Research Systems SRS 830 lock-in amplifier, and the read-out signal was fed to the JPK scanning probe module (SPM) control for data sampling and storage. A calibrated photodetector was used to measure the transmitted light through the device. The transmission map of the device is used to correlate the local photocurrent value to discern the position of the metal electrodes. The ambient light was completely blocked to avoid any external exposure to the sample. The measurements were performed in ambient air due to the practical limitations of mounting the instrument in a vacuum. During the measurements, desiccants were placed around the setup to reduce the moisture content in the ambient atmosphere.

For the spatially-resolved NSPM measurement, in addition to the measurements on long channel devices, we studied the profiles for a short channel device of $l = 13$ μm.

*Transient Photocurrent*

Transient Photocurrent measurements were carried out by using a pulsed supercontinuum laser (YSL Supercontinuum SC PRO 7, $\lambda_{ex}$ ~ 400-2300 nm) with a pulse width of ~ 100 ps (with 100 kHz repetition rate) and energy ~ 1 μJ pulse$^{-1}$, incident from the glass-substrate side on the lateral device. The device was operated at 5 V bias, and the current was measured in series using a digital oscilloscope (Tektronix MDO3024, 2.5 GS s$^{-1}$) with 50 Ω input coupling.

**References:**

[1] N. Mott, GURNEY RW-Electronic processes in ionic crystals (Oxford University Press, 1940).

[2] M. A. Lampert and P. Mark, *Current injection in solids* (Academic press, 1970).

[3] J. A. Röhr and R. C. MacKenzie, Analytical description of mixed ohmic and space-charge-limited conduction in single-carrier devices, **Journal of Applied Physics 128, 165701 (2020)**.

[4] O. E. Semonin, G. A. Elbaz, D. B. Straus, T. D. Hull, D. W. Paley, A. M. Van der Zande, J. C. Hone, I. Kymissis, C. R. Kagan, and X. Roy, Limits of carrier diffusion in n-type and p-type CH3NH3PbI3 perovskite single crystals, **The journal of physical chemistry letters 7, 3510 (2016)**.

[5] Q. Dong, Y. Fang, Y. Shao, P. Mulligan, J. Qiu, L. Cao, and J. Huang, Electron-hole diffusion lengths > 175 μm in solution-grown CH3NH3PbI3 single crystals, **Science 347, 967 (2015)**.